\documentclass[twocolumn]{aastex631}

\usepackage{multirow}
\usepackage{amsmath}

\shorttitle{Paper II}
\shortauthors{Balakrishnan et al.}

\newcommand{\Chandra}{{\sl Chandra}}

\newcommand{\SgrA}{Sgr~A*}
\newcommand{\FeKalpha}{Fe~K$\alpha$}
\newcommand{\vlos}{$v_{\rm LoS}$}

\begin{document}

\title{Multistructured accretion flow of Sgr A* II: Signatures of a Cool Accretion Disk in Hydrodynamic Simulations of Stellar Winds}

\correspondingauthor{Mayura Balakrishnan}
\email{bmayura@umich.edu}

\author[0000-0001-9641-6550]{Mayura Balakrishnan}
\affil{Department of Astronomy, University of Michigan, 1085 S. University, Ann Arbor, MI 48109, USA}

\author[0000-0002-9213-0763]{Christopher M.\ P.\ Russell}
\affiliation{Department of Physics and Astronomy, Bartol Research Institute, University of Delaware, Newark, DE 19716, USA}

\author[0000-0002-5466-3817]{Lia Corrales}
\affil{Department of Astronomy, University of Michigan, 1085 S. University, Ann Arbor, MI 48109, USA}

\author[0000-0002-9019-9951]{Diego Calder\'on}
\affiliation{Hamburger Sternwarte, Universität Hamburg, Gojenbergsweg 112, D-21029 Hamburg, Germany}

\author[0000-0003-1965-3346]{Jorge Cuadra}
\affiliation{Faculty of Liberal Arts, Universidad Adolfo Ibañez, Av. Padre Hurtado 750, Viña del Mar, Chile}
\affiliation{Millennium Nucleus on Transversal Research and Technology to Explore Supermassive Black Holes (TITANS)}

\author[0000-0001-6803-2138]{Daryl Haggard}
\affiliation{Department of Physics, McGill University, 3550 University Street \#040, Montreal, QC H3A 2A7, Canada}

\author[0000-0001-9564-0876]{Sera Markoff}
\affiliation{Anton Pannekoek Institute for Astronomy/GRAPPA, University of Amsterdam, Science Park 904, rm. C4.151
1098 XH Amsterdam, Netherlands}

\author[0000-0002-8247-786X]{Joey Neilsen}
\affiliation{Department of Physics, Villanova University, 800 Lancaster Avenue, Villanova, PA 19085, USA}

\author[0000-0001-6923-1315]{Michael Nowak}
\affiliation{Department of Physics, Washington University at St. Louis, 1 Brookings Dr, St. Louis, 63130 MO, USA}

\author[0000-0002-9279-4041]{Q.\ Daniel Wang}
\affiliation{Department of Astronomy, University of Massachussets Amherst, 710 North Pleasant Street Amherst, MA 01003, USA}

\author[0000-0003-3852-6545]{Frederick Baganoff}
\affiliation{MIT Kavli Institute for Astrophysics and Space Science, MIT, 70 Vassar St, Cambridge, MA 02139, USA}



%% Mark off the abstract in the ``abstract'' envIronment. 
\begin{abstract}

Hydrodynamic simulations of the stellar winds from Wolf-Rayet stars within the Galactic Center can provide predictions for the X-ray spectrum of supermassive black hole \SgrA. Herein, we present results from updated smooth particle hydrodynamics simulations, building on the architecture of \citet{Cuadra2015, Russell2017}, finding that a ``cold'' ($10^4$ K) gas disk forms around \SgrA\ with a simulation runtime of 3500 years. This result is consistent with previous grid-based simulations, demonstrating that a cold disk can form regardless of numerical method. We examine the plasma scenarios arising from an environment with and without this cold disk, by generating synthetic spectra for comparison to the quiescent \FeKalpha\ \SgrA\ spectrum from \Chandra\ HETG-S, taken through the \Chandra\ X-ray Visionary Program. We find that current and future X-ray missions are unlikely to distinguish between the kinematic signatures in the plasma in these two scenarios. Nonetheless, the stellar wind plasma model presents a good fit to the dispersed \Chandra\ spectra within 1.5\arcsec\ of \SgrA. We compare our results to the Radiatively Inefficient Accretion Flow (RIAF) model fit to the HETG-S spectrum presented in Paper I and find that the Bayesian model evidence does not strongly favor either model. With 9\arcsec\ angular resolution and high spectral resolution of the X-IFU, \textit{NewAthena} will offer a clearer differentiation between the RIAF plasma model and hydrodynamic simulations, but only a future X-ray mission with arcsecond resolution will significantly advance our understanding of \SgrA's accretion flow in X-rays.

\end{abstract}



%%======================================
\section{Introduction} \label{sec:intro}
%%======================================

Sagittarius A* (\SgrA), the supermassive black hole in the center of the Milky Way, is immersed in a complex, multi-phase gaseous medium, and surrounded by a parsec-scale nuclear star cluster \citep[see][for a review]{Genzel2010review}. While most of the cluster is comprised of old stars \citep{Nogueras2020}, there is also a young population, including some 30 Wolf-Rayet (WR) stars with orbits that extend as close as $\approx 0.05$ pc from \SgrA \citep{Paumard2006, Lu2013}. WR stars are evolved, massive stars, characterised by powerful stellar winds that have mass-loss rates $\dot M \sim 10^{-5} M_\odot\,$yr$^{-1}$ and terminal speeds $v \approx 500 - 2500\,$km~s$^{-1}$ \citep{Martins2007}. The relatively small separation between the WR stars means that their stellar winds collide in relatively high-density shocks, producing plasma that reaches $10^7$ K, emits in X-ray \citep{Baganoff2003, quataert2004, Wang2013}, and is accreted onto the supermassive black hole \citep{Cuadra2005}. 

The number of relevant wind sources and their non-trivial orbital distribution \citep{vonFellenberg2018, Jia2023}, makes this region hard to model analytically, in particular if we want to obtain the amount of gas available for accretion onto \SgrA. Several studies have therefore relied on hydrodynamic simulations to model the gas dynamics of the material supplied by the WR stars orbiting \SgrA. The predicted accretion rate arising from hydrodynamic simulations \citep[][hereafter C15]{Cuadra2015} at the Bondi radius ($\sim 4-5$\arcsec, or 0.2 pc; $\dot M_{\rm SMBH} \sim 10^{-6} M_\odot\,$yr$^{-1}$) is in agreement with theoretical estimates based on X-ray observations at the same scale \citep{Baganoff2003}. Further numerical work has reached deeper into the Galactic center,  by first running parsec-scale simulations of the stellar winds \citep{Ressler2018,Ressler2020MHD}, and then zoom-in magnetohydrodynamic (MHD) and general relativistic magnetohydrodynamic (GRMHD) models of regions progressively closer to \SgrA~\citep{Ressler2020GRMHD}. While these simulations do not currently fit observations close to the event horizon \citep{EHT2022V}, they provide a promising framework for understanding how the observed properties of the accretion flow came to be.

The simulations used in C15, \citet[][hereafter R17]{Russell2017}, and \citet{Calderon2020} focus on the WR stars as they have the largest mass-loss rates of any stars in the region \citep{Martins2007}. There is likely some outflow produced by the accretion flow of Sgr A* itself, predicted by simulations \citep{Ressler2018, Dexter2020, Chatterjee2021} and Radiatively Inefficient Accretion Flow (RIAF) models \citep[e.g.,][]{BB1999}. RIAFs are parameterized by mass accretion rate, $\dot{m} \propto r^s$, material density, $n \propto r^{-3/2+s}$, and temperature $T \propto r^{-q}$. Here, $s = 0$ corresponds to classical Bondi accretion and $s = 1$ indicates an inflow balanced by an outflow \citep{Begelman2012}. Observational fits to the multiwavelength spectral energy distribution (SED) of Sgr A* and \Chandra\ ACIS spectrum with RIAF models suggest varying $s$ values, with $s \sim 0.05-0.3$ closer to the BH \citep{Yuan2003, Ma2019} and $s \sim 0.6 - 1$ at the Bondi radius \citep{Ma2019, Wang2013}, demonstrating that this outflow is more prominent at higher radii. However, incorporating this outflow into WR simulations has minimal impact on the predicted X-ray spectral shape and only marginally reduces the X-ray flux compared to simulations without \SgrA\ feedback (R17). Main sequence OB-star winds, with mass-loss rates three orders of magnitude smaller than the WR stars, may also affect the net mass accretion rate in this region but are found to underpredict the accretion rate at the Bondi radius \citep{Luetzgendorf2016}. Therefore, we solely consider mass loss from WR stars for our analysis. 

The WR stars alone are able to explain an exceptional amount of the observed behaviour in this region, namely the predicted emission fully explains the X-ray spectrum of \SgrA\ (R17). R17 post-processed the simulations of C15 using radiative-transfer tools, producing synthetic X-ray maps and spectra.  Analysis of the synthetic observations shows that the overall X-ray flux depends mostly on the mass-loss rates, while the spectral shape is determined by the wind velocities. R17 found that the models closely reproduce the 2-10 keV \Chandra\ ACIS spectrum in the 2\arcsec$-$5\arcsec\ region. This is rather noteworthy, as the models take as input the stellar wind and orbital properties obtained from infrared observations \citep{Paumard2006, Martins2007}, and are therefore completely independent of the X-ray data they explain. 

The stellar wind simulations predict other observed structures. Depending on the stellar wind collision properties, the plasma can be prone to instabilities and form dusty, gaseous blobs \citep{Fritz2011,Peissker2023}, which may explain some of the gaseous structure observed in infrared and radio in the same region \citep{Gillessen2012}.  Moreover, the stellar winds can perturb some of the larger-scale gaseous streams reaching the central region \citep{Wang2020} or the circumnuclear disk \citep{Solanki2023}. The mechanisms that enable cold and hot phase plasma structures to co-exist in this way remains a mystery.

There have been claims of detection of a cool ($\sim 10^4$ K) H30$\alpha$ disk co-spatial with the $\sim 10^7$ K accretion flow. \citet{murchikova2019} measured line shifts at 231.9 GHz with the Atacama Large Millimeter/submillimeter Array (ALMA), finding red- and blue-shifted components approximately 0.004 pc from \SgrA. The estimated mass of this ionized disk structure is approximately $10^{-4} - 10^{-5} M_\odot$.  These results were confirmed in \citet{Yusef-Zadeh2020}, who also found these line shifts in H39$\alpha$, H52$\alpha$, and H56$\alpha$ emission. However, \citet{Ciurlo2021} obtained observational upper limits for the undetected Br$\gamma$ disk ($\sim 6000$ K) that are 80 times lower than the expected emission given the H30$\alpha$ flux. These authors also argue that a maser explanation for this discrepancy, proposed originally by \citet{murchikova2019}, is not consistent with the expected properties of the plasma nor with the putative disc geometry. The results in \citet{Ciurlo2021}, however, are consistent with Br$\gamma$ emission levels predicted by \citet{Calderon2020}, who showed that running the simulations for long-enough timescales ($>$ 3000 yr) leads to the accumulation of a $\approx 0.01$~pc cold gas disk ($\sim 10^4$ K) around \SgrA. This study utilized \texttt{ramses}, which follows an Eulerian approach and incorporates adaptive mesh refinement (AMR) for increased resolution, identifying two distinct mass accretion phases: quasi-steady and disk regimes. The quasi-steady phase, characterized by $\dot M_{in} > \dot M_{out}$, eventually transitions into the disk phase when a bow shock facilitates rapid cooling, leading to the formation of a cold gas disk. This process results in increased rates of inflow and outflow, eventually reaching a balance where $\dot M_{in} \sim \dot M_{out}$, indicative of $s \sim 1$ as predicted by RIAF models. We note that this disk is not generated in another series of simulations \citep{Ressler2018, Ressler2020MHD}, even after extending the simulation run time to 9000 yrs in the past \citep{Ressler2020MHD}. This discrepancy could be due to differences in the cooling prescriptions; the effect of cooling on the formation of this disk is investigated further in Calderón et al. 2024 (in prep). 

To test whether the cold disk formed in \citet{Calderon2020} is an artifact of the grid-based approach, we extend the SPH simulations of R17 to a longer timescale of 3500~yr. In addition, we test whether we can identify signatures of the cool accretion disk in the particle-based simulations by examining the predicted X-ray emission in both plasma scenarios.
%%%%%%%%%
%
Although the predicted ``cold'' ($\sim$10$^4$ K) disc is not directly observable in X-rays, we wish to investigate whether its kinematics might leave a signature in the hot gas.  \Chandra\ HETG currently offers the best available spectral resolution combined with sub-arcsecond imaging resolutions, which is required to study the \SgrA\ accretion flow. While the low-resolution spectrum from this dataset is reproduced by the stellar winds (R17), we leverage the legacy \Chandra\ HETG-S dataset from \citet{Corrales2020} to test whether hydrodynamic simulations of shock-heated WR stellar winds are also able to reproduce the observed X-ray spectrum at the high spectral resolution offered by \Chandra. The High Energy Transmission Grating Spectrometer \citep[][HETGS]{CanizaresHETG} on \Chandra\ disperses a fraction of the incoming photons, providing a high-resolution spectrum with a spectral resolution between 60$-$1000. It consists of two gold gratings, MEG (0.4$-$5 keV sensitive) and HEG (0.8$-$10 keV sensitive, $E/dE \sim 170$ at 6.7 keV). Here we focus on the \FeKalpha \ complex ($\sim$ 6.7 keV) of the \SgrA\ accretion flow captured by the HEG \citep{Corrales2020} to evaluate the possibility that the observed line structure is a direct probe of the velocity structure in the accretion flow.

In Section \ref{sec:methods}, we describe our methodology, with details on the hydrodynamic simulation of the Wolf-Rayet winds followed by how predicted HEG spectra are generated. In Section \ref{sec:results}, we show our X-ray line profiles and results from fitting these predictions to the HEG data. We compare the stellar wind profiles to the best fits for a RIAF model, presented in a companion paper (Balakrishnan et al. 2024a, referred to hereafter as Paper I), and show micro-calorimeter predictions of these two models. We end with a summary of our results and conclusions.

%%===================================
\section{Methods} \label{sec:methods}
%%===================================

The quiescent high-resolution spectrum of \SgrA\ was extracted from the 3~Ms \Chandra\ (PIs: Markoff, Nowak, \& Baganoff) campaign in 2012 in \citet{Corrales2020}. 
%\citet{Corrales2020} removed background and Sgr A* flares, resulting in 2.55 Ms of observations. 
After removing flares from the soft particle background and from Sgr~A* itself, the exposure time totals to 2.55~Ms. 
The counts spectrum was created by using customized background extraction regions that exclude the nearby sources pulsar wind nebula PWN~G359.945-0.045 and star cluster IRS~13E. This work focuses on the \FeKalpha\ complex, which is the strongest X-ray line emission feature from \SgrA, and we accordingly utilize the spectrum only between 6.4 keV and 7.2 keV. See \citet{Corrales2020} for further details on the cleaning and production of the spectrum. 

\begin{figure}
    \centering
    \includegraphics[width=0.48\textwidth]{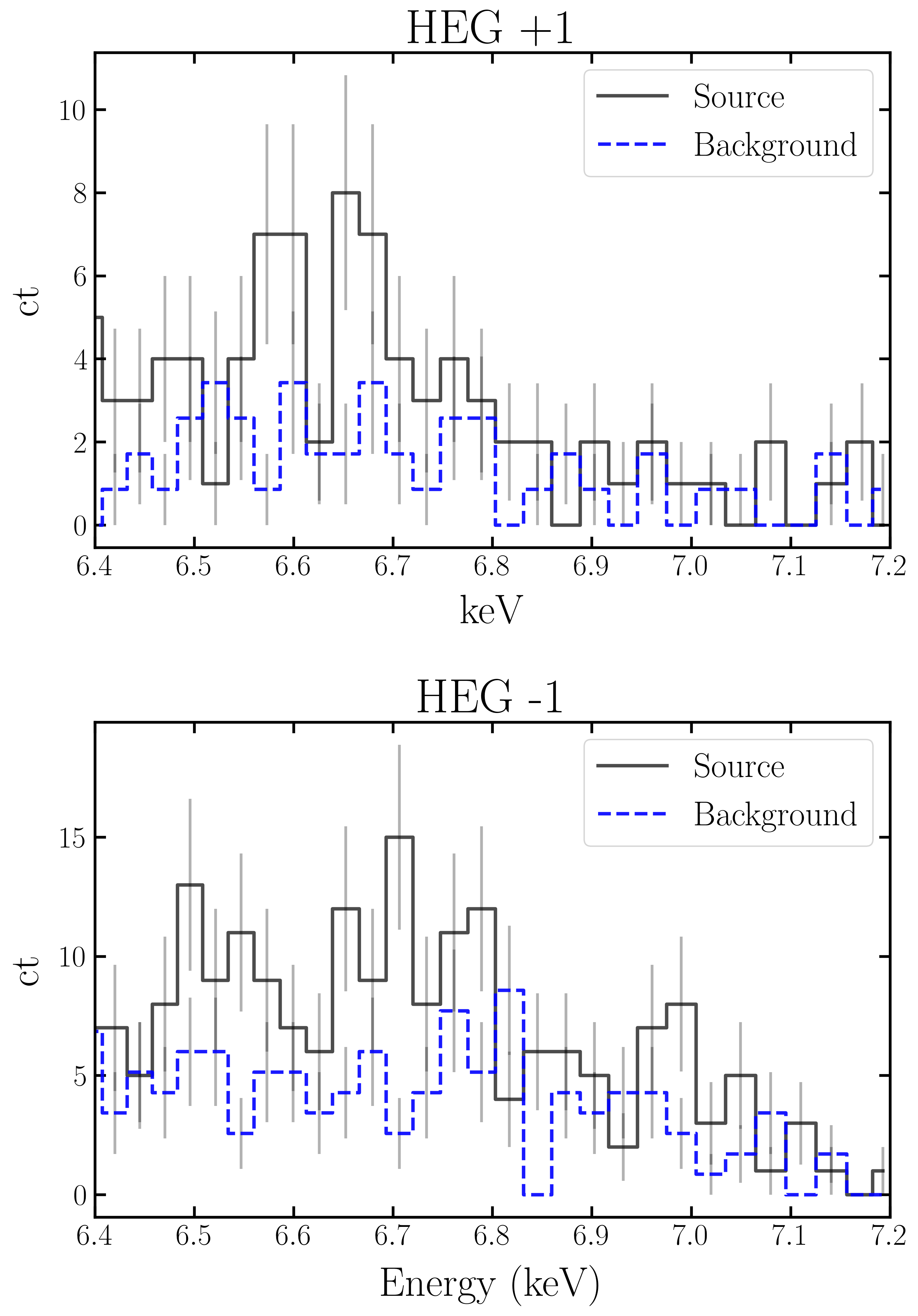}
    \caption{The high-resolution quiescent spectrum of \SgrA, extracted in \citet{Corrales2020}, plotted in counts. The top and bottom panels show the data from the HEG +1 and -1 orders, respectively. In this work, we focus on the \FeKalpha\ complex, between 6.4 and 7.2 keV. The spectra extracted from the source (out to 1.5 \arcsec) and background (between 1.5\arcsec and 5\arcsec) regions are plotted in black and blue, respectively, and are rebinned to a factor of 3.}
    \label{fig:data}
\end{figure}

Figure~\ref{fig:data} shows the counts histograms for the source (0$-$1.5\arcsec; in solid black) and background (1.5\arcsec$-$5\arcsec; in dashed blue) regions, for both the HEG +1 and -1 orders. The data are binned to a factor of 3 for visual clarity. Peaks around 6.7 keV are visible in the source regions; numerous other features suggest potential velocity or temperature variations, but might also arise from statistical noise. The velocity resolution of the Chandra HEG is $\Delta v \sim 1500 \text{km} \text{ s}^{-1}$, which is close to the maximum line-of-sight velocities seen in the stellar wind simulations, indicating we may be able to probe the complex velocity structure in the \SgrA\ accretion flow with the high resolution spectrum.
%To compare these observations with theoretical predictions, we utilize simulations from \citet{Cuadra2015} as analyzed in R17. 
To examine whether these line features can be explained by a non-homogeneous, multi-temperature plasma with velocity features, we utilize updated simulations from C15 and R17. We calculate the expected X-ray emission generated by stellar winds for each of the source and background regions, and process this thermal emission, accounting for any effects induced by the HETG-S detector. See Paper I for more details regarding corrections made for HETG-S geometry.

%%============================================
\subsection{Simulation Setup} \label{sec:sims}
%%============================================

% 
This work utilizes simulations for the plasma environment of \SgrA\ created by WR stars in the central parsec, as developed by C15 and R17. These simulations use the smoothed particle hydrodynamics (SPH) code $\texttt{GADGET-2}$ \citep{Springel2005} to follow the orbits of the 30 WR stars around Sgr~A*, all while ejecting their stellar winds into the simulation domain of a half parsec in radius. The WR stellar mass-loss rates, wind speeds, and spectral types come from \citet{Martins2007}, and the stellar orbits are from \citet{Paumard2006}. These supersonic winds collide with one another, creating a complex structure of cold, free-streaming winds and hot, post-shock gas that emits thermal X-rays.  Material flowing towards the SMBH is removed from the simulation at an inner radius of $r_{\rm min}=1.2\times10^{16}$cm~$\approx 10^4 R_g$ (where $R_g = GM_*/c^2$), corresponding to a projected angular distance of 0.02\arcsec~from \SgrA\, while material flowing outward is removed at $r_{\rm max}=1.2\times10^{18}$cm~$\approx 2.4 \times 10^6 R_{g}$, corresponding to a projected angular distance of $12\arcsec$.

In order to reproduce the work of \citet{Calderon2020},  which found two different accretion modes onto \SgrA (quasi-steady and disk mode), we perform two sets of simulations that have different initial starting times. Starting from 3500~yr \citep[as done in ][]{Calderon2020} in the past yields a cold disk around the SMBH, while starting from 1100~yr ago (as done in C15) does not. Additionally, compared to the C15 simulations, the SMBH mass has been updated to $M_{\rm SMBH}=4.28\times10^{6}M_\odot$ \citep{Gillessen2017}, and the separation of the WR stars in the IRS~13E cluster has been increased to bring the cluster's X-ray flux into better agreement with the observations \citep[R17; see also][]{Wang2020}. We note that this mass is slightly different to that used in Paper I \citep[$M_{\rm SMBH}=4.15\times10^{6}M_\odot$;][]{gravity2019}; however, changing the mass in the RIAF model to match \citet{Gillessen2017} changes the counts histogram by less than 4\%. These versions of the simulations do not take feedback from the SMBH into account; as shown in R17, this feedback does not change the shape of the X-ray spectrum significantly. The simulations include the same radiative cooling prescription as C15, which uses a $Z=3Z_\odot$ cooling curve for all winds. 

We can predict the HEG spectra from both sets of simulations to see if we can distinguish between the two accretion modes with X-ray spectra. A model spectrum for the thermal X-ray emission is synthesized from the hydrodynamic simulations in a similar fashion as R17, which is built upon a modified version of the visualization program \texttt{Splash} \citep{Price2007}.  All mass elements hot enough to emit thermal X-ray emission over $\sim 10^6$ K (indicated by superscript $k$) are assigned energy-dependent X-ray emissivities,
$j_E^k=n_e^kn_i^k\Lambda(E,T^k)$ from the \texttt{vvapec} model \citep{Smith2001} \citep[obtained from \texttt{XSPEC},][]{Arnaud1996}.  The temperature and emission measure of each mass elements' \texttt{vvapec} spectrum $\Lambda(E,T^k)$ corresponds to the temperature ($T^k$) and density (electron number density $n_e^k$ and ion number density $n_i^k$) of the mass element computed in the hydrodynamic simulation. To distinguish between the various abundances of the winds, the winds and their \texttt{vvapec} spectra are split into three categories: WC8-9 stars, WN6-7 stars, and WN8-9 \& Ofpe/WN9 stars (following R17). The optical depths through the simulation domain ($\int\kappa_E\rho^k dz$ with $z$ being the line-of-sight integration direction) in the observed X-ray energy range are in the optically thin limit; this calculation made use of X-ray opacities ($\kappa_E$) of \citet{VY1995} %\citet{Verner1995} 
obtained via an abundance-modified version of the \texttt{windtabs} model \citep{Leutenegger2010}. 

While the density and temperature of the plasma is set by the hydrodynamic simulations, which use a $3~Z_\odot$ cooling prescription, our predicted Iron line profiles were calculated with abundances of $Z_{Fe} = 0.6-1~Z_\odot$ \citep{Asplund2009} depending on the specific WR subtype (see Table 1 of R17 for details). A high metal abundance in the plasma is well-justified; WR stars have low (or non-existent) Hydrogen abundances but high Carbon, Nitrogen, or Oxygen abundances \citep{Sander2019}. Being young stars, WR stars are indicators of recent star formation, so it is reasonable to expect an Iron abundance $\sim Z_\odot$ or higher in the X-ray emitting plasma due to past stellar explosions. However, X-ray studies of WR stars are rare, and there is not yet a generally accepted value. Recent spectral fitting of the 6.7 keV line in the S~308 bubble around a galactic WR system finds a Nitrogen abundance of $\sim 3~Z_\odot$ and an Iron abundance consistent with $1~Z_\odot$ \citep{Camilloni2024}. Other studies have used Iron abundances as low of $0.4~Z_\odot$ \citep{Sander2019}. These abundances are similar to those found from fits to the X-ray emitting plasma around \SgrA. The RIAF fit from \citet{Wang2013} found an Iron abundance of $Z_{Fe} = 1.23~Z_\odot$ \citep{Asplund2009}, while in Paper I, we found that the spectrum was best-fit an average Iron abundance of $Z_{Fe,\rm \ Paper~I} = 0.18~Z_\odot$ \citep[relative to][; see Section \ref{sec:results_b} for further discussion]{Asplund2009}.

We calculate the thermal X-ray intensity from the \SgrA\ plasma with the X-ray emissivities above, while incorporating the line-of-sight velocities ($v_{\rm LoS}$) in each plasma element. The profile function for mass element $k$ is a Gaussian of the form:
\begin{equation}
\phi^k(v_{\rm LoS})= \frac{1}{\sqrt{\pi}} \exp\left(\left(v_{\rm LoS}^k-v_{\rm LoS}\right)^2/v_{\rm th}^2\right)
\end{equation}
where $v_{\rm LoS}^k$ is the element's line-of-sight velocity, $v_{\rm LoS}$ is the particular line of sight velocity being computed, and $v_{\rm th}$ is the thermal/sound speed. \texttt{Splash} sums the contribution of each mass element across the $\{x,y\}$ pixel map to obtain thermal X-ray intensity at a particular energy, and we incorporate $v_{\rm LoS}$ as follows:
\begin{equation}\label{eq:intensity}
I_E(x,y) = \int j_E^k (x,y) dz = \int n_e^k n_i^k \Lambda(E,T^k) \phi^k(v_{\rm LoS}) dz \ 
\end{equation}
where we integrate $z$, or line-of-sight direction, from $-2.4 \times 10^6 R_g$ to $2.4 \times 10^6 R_g$ where \SgrA\ is situated in the plane of $z=0$. Computing a single energy's line profile requires looping over the full range of \vlos, which ranges from -3000 to 3000 km s$^{-1}$ in the simulation. The Fe complex has many lines that overlap each other when their $v_{\rm LoS}$ ranges are converted to energy space, so we compute a line profile for each entry in the VVAPEC tables, yielding 150 line profiles at an energy resolution of 6400/dex.  All of these overlapping line profiles are then finely interpolated to the same energy grid, where they are summed to yield the complete $v_{\rm LoS}$-dependent spectrum in the Fe complex. Similarly, in the continuum region, calculations at specific energies blend into adjacent lines due to their proximity, resulting in a composite spectrum. Therefore, we conduct comprehensive line profile calculations across every point within both the continuum and line-complex regions to account for their influence on one another.

\begin{figure*}[ht!]
    \centering
    \includegraphics[height=4.55cm,clip,trim={0 4.3cm 0 0}]{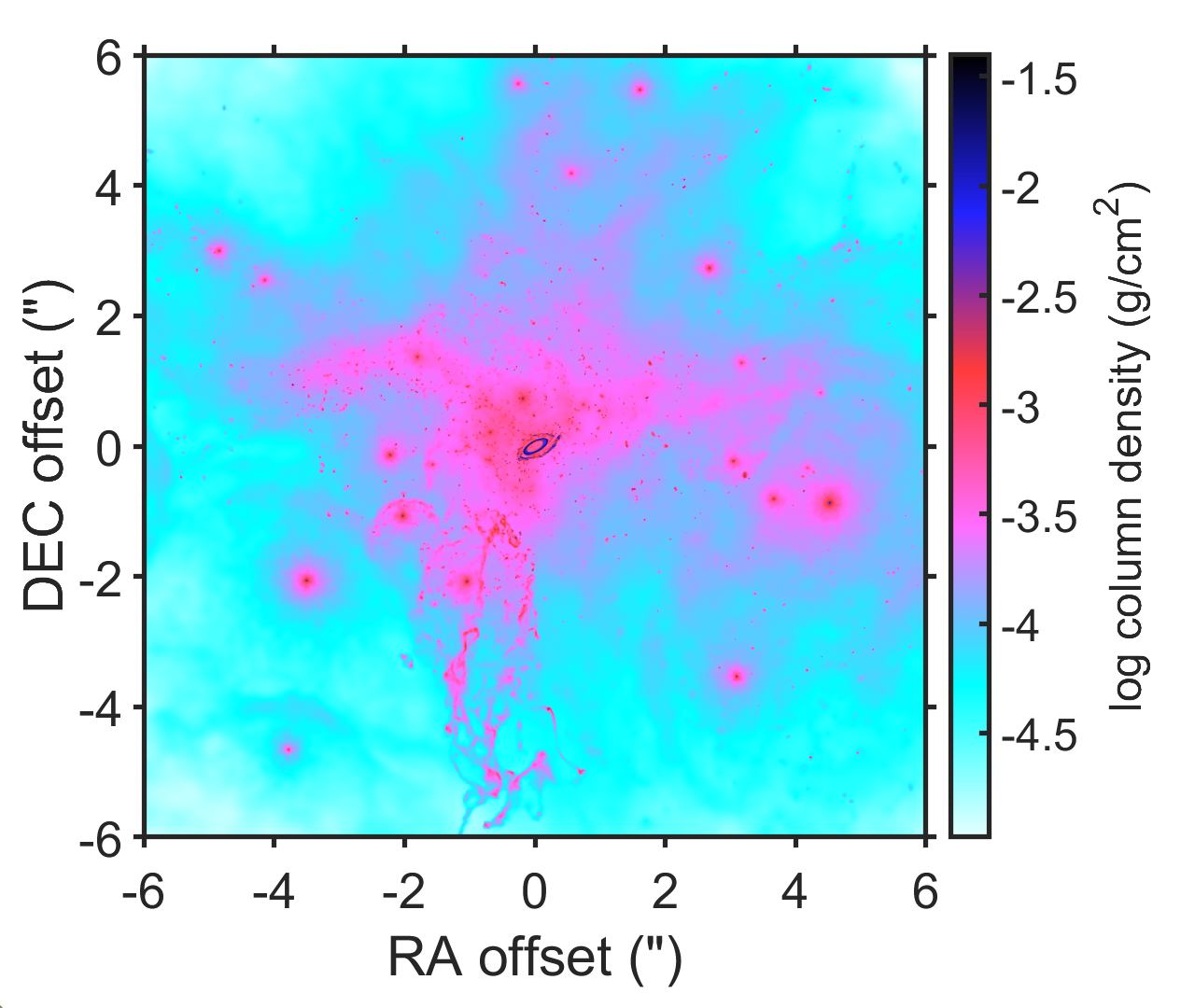}
    \includegraphics[height=4.55cm,clip,trim={3.5cm 4.3cm 0 0}]{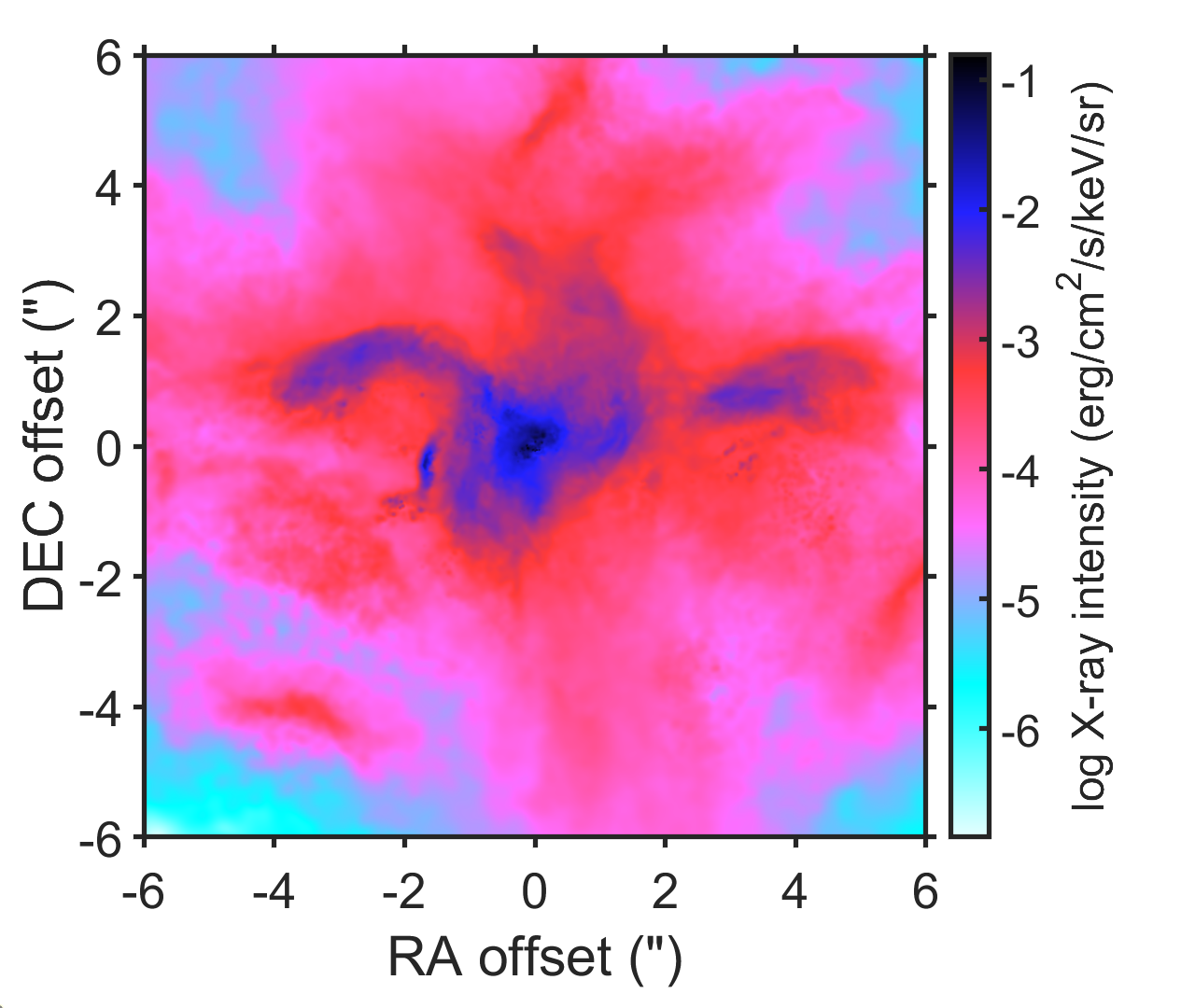}
    \includegraphics[height=4.55cm,clip,trim={3.5cm 4.3cm 0 0}]{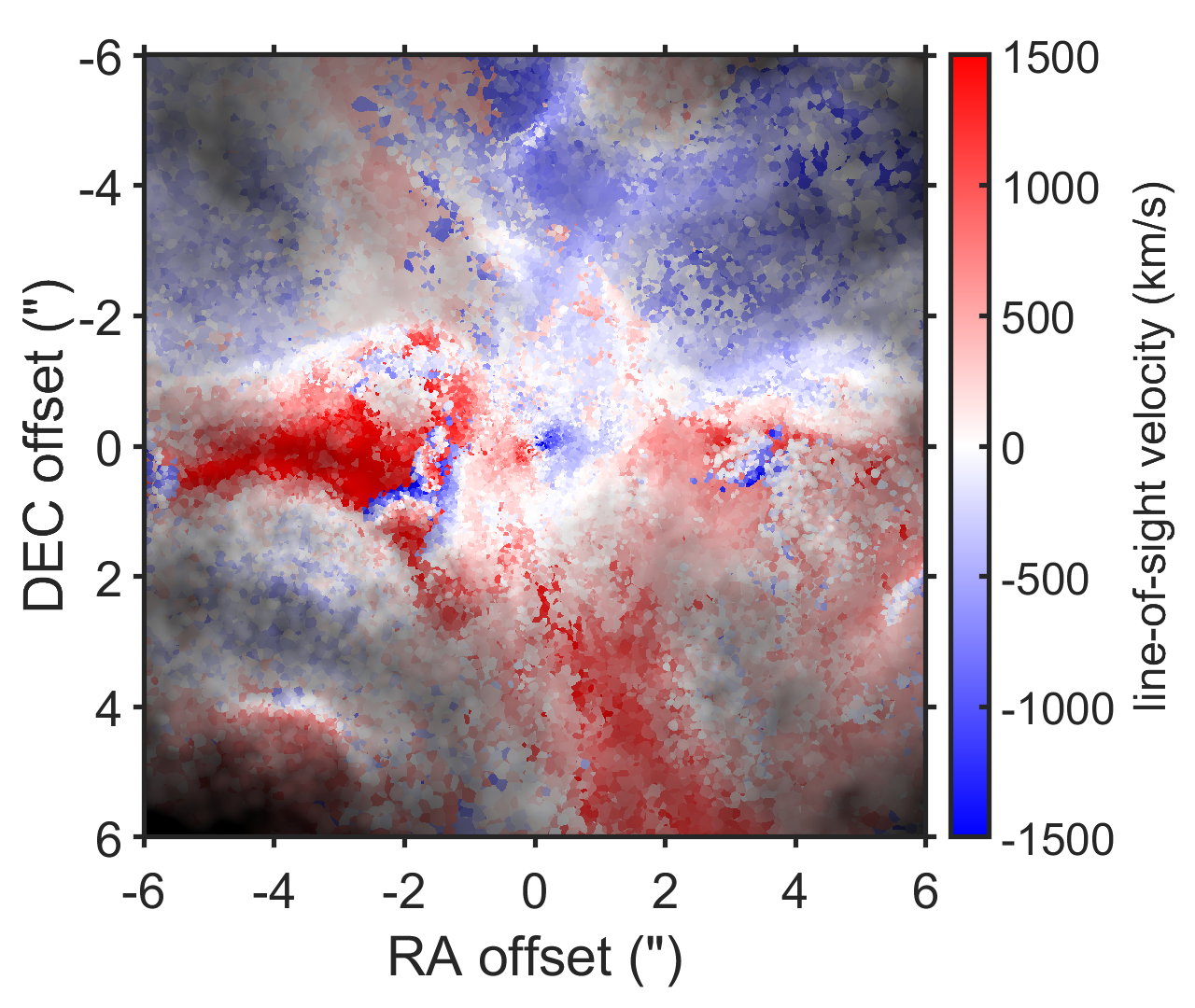}%
    
    \includegraphics[height=5.35cm]{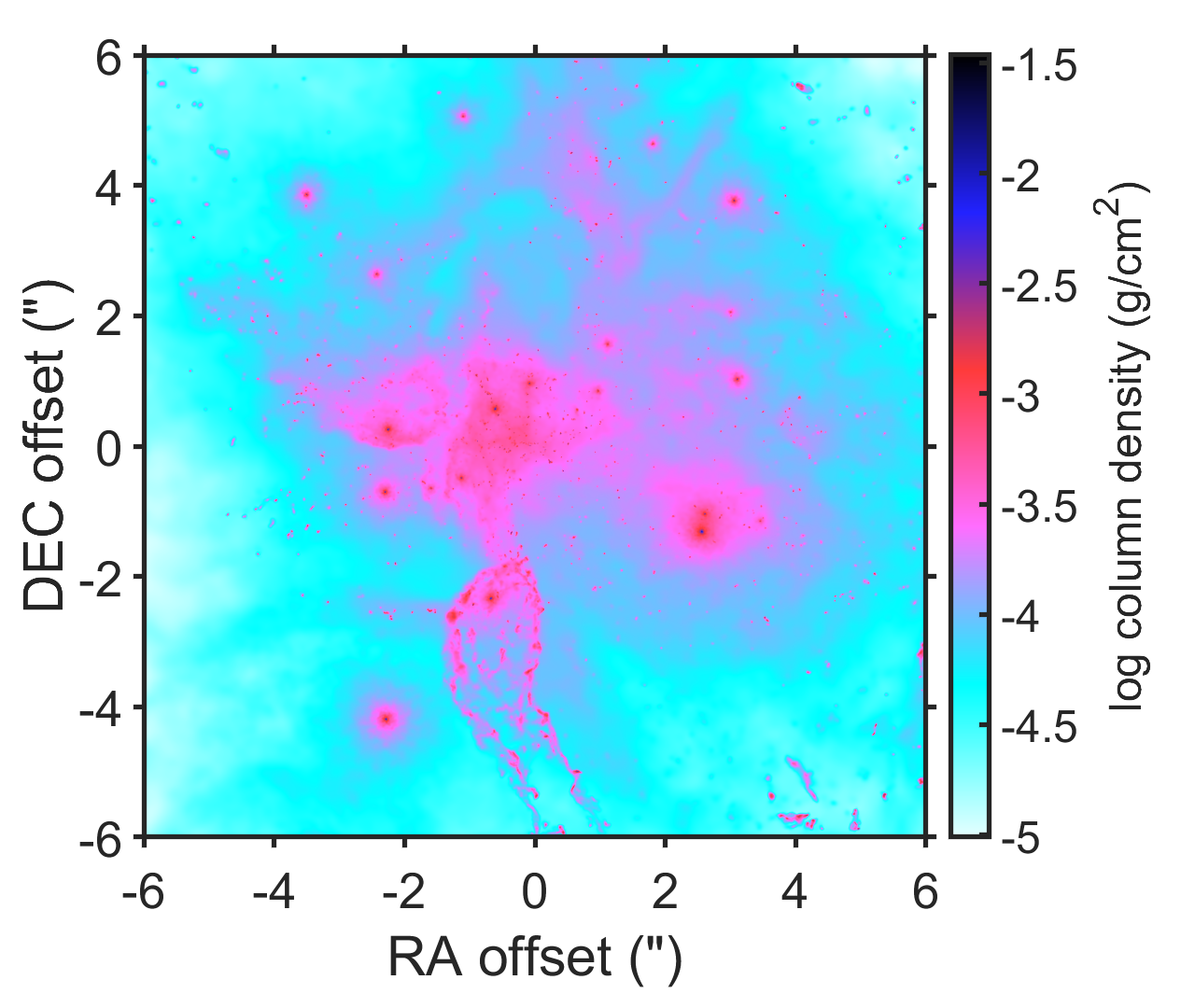}
    \includegraphics[height=5.35cm,clip,trim={3.5cm 0 0 0}]{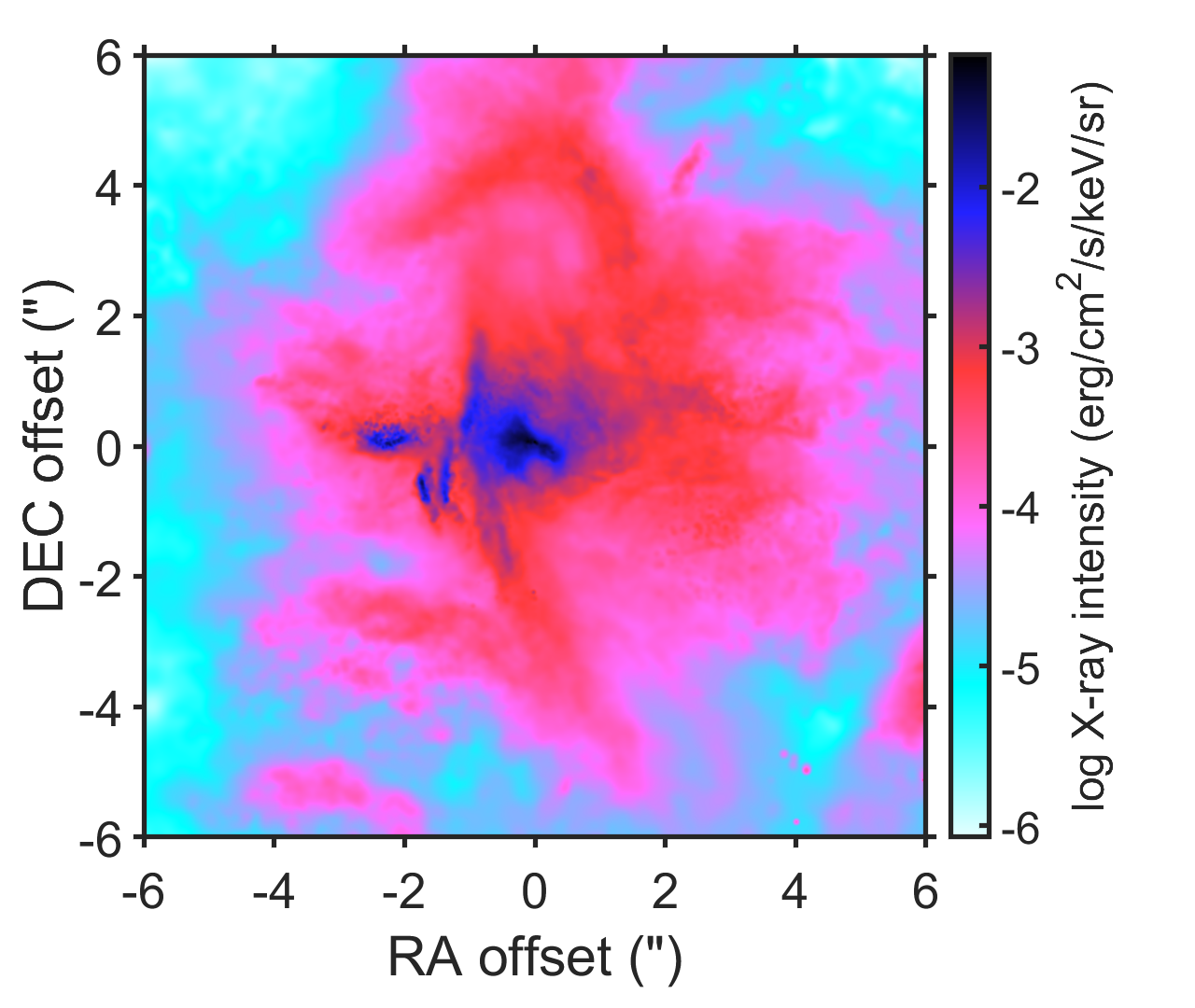}
    \includegraphics[height=5.35cm,clip,trim={3.5cm 0 0 0}]{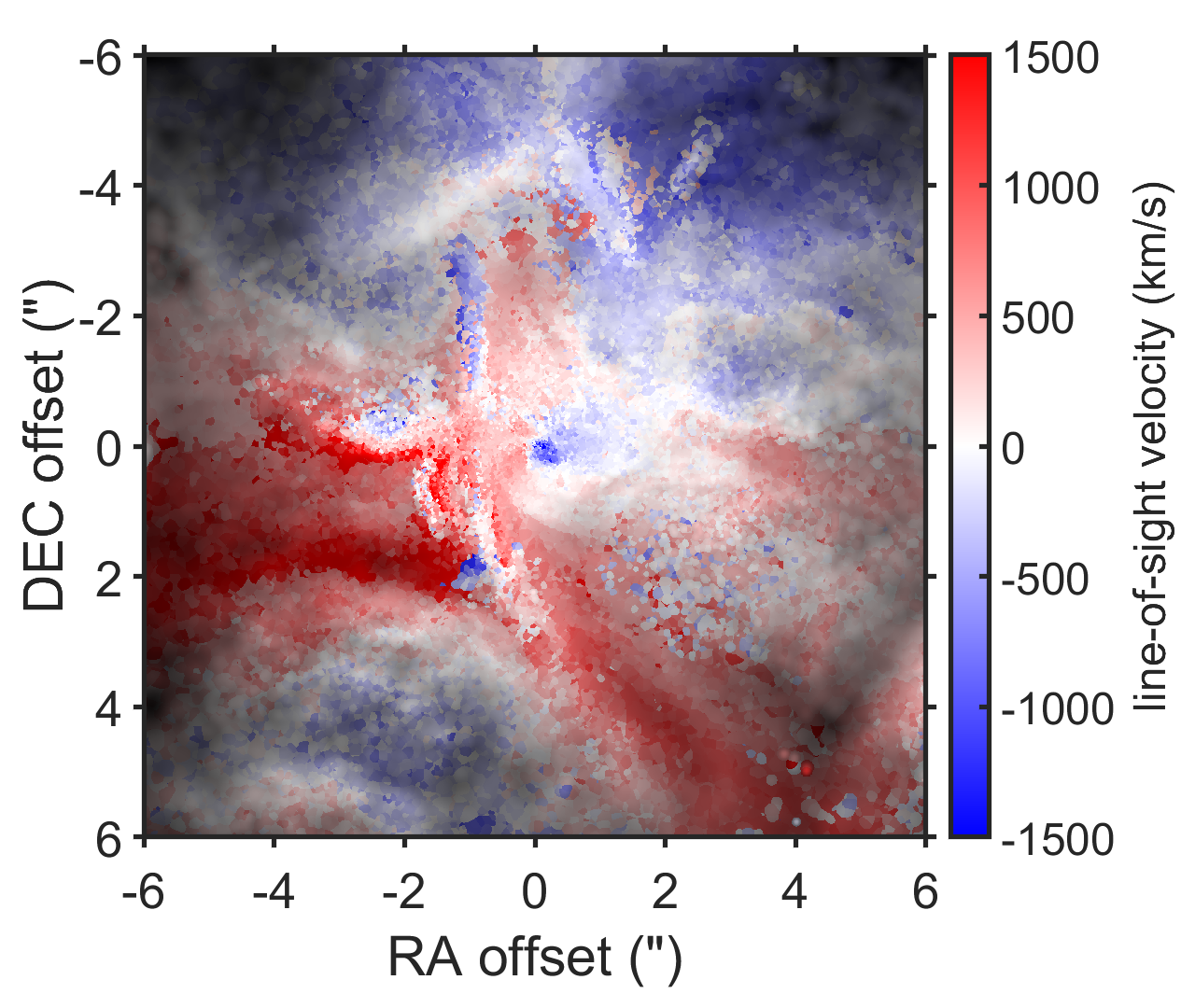}
    \caption{The top and bottom rows show the results of the disk (runtime of 3500 yr) and no-disk (runtime of 1100 yr) scenarios, respectively, for gas emitting at $6.7$~keV. The plots span the central $12 \arcsec \times 12 \arcsec$ surrounding \SgrA, with North upwards and East to the left. LEFT: Column density map of the region, with the cool disk of gas seen clearly in the top left panel. MIDDLE: 6.7 keV X-ray intensity with the dark blue highlighting high levels of \FeKalpha \ emission. RIGHT: Illustration of the velocity structure in this complex plasma, with color denoting the line-of-sight velocity corresponding to the line profile's maximum X-ray brightness for each pixel. The grayscale overlay reflects X-ray emission intensity, i.e. dimmer regions indicate X-ray dim gas. }
    \label{fig:sims}
\end{figure*}

In Figure \ref{fig:sims}, we show the column density, X-ray intensity, and velocity maps corresponding to the disk (top row) and no-disk plasma scenarios (bottom row). Large scale density and X-ray emissivity field differences are at least in part due to the different run times, with the top row showing a snapshot at $t = 3500$ yr and the bottom row reflecting the plasma conditions at $t = 1100$ yr. The disk is visible in the top left plot owing to its high column density. Contributions from slower winds result in high density shocks that cool radiatively, resulting in instabilities that lead to clumpy structure seen in the images. To make the velocity maps, we split the X-ray emission of the same line into 5 km/s bins for each pixel's \vlos. Subsequently, we assigned colors to each pixel based on the \vlos\ corresponding to the maximum velocity-binned X-ray emission, utilizing a blue-white-red color gradient. Next, to highlight X-ray bright areas and consequently identify X-ray dim regions, we combined this color-coded \vlos\ image with a grayscale image representing X-ray brightness. This process preserves the X-ray bright sections of the velocity map while enhancing the contrast of the X-ray dim areas. The disk and quasi-steady accretion scenarios lead to different kinematics in the gas that might be resolvable with the high-resolution \Chandra\ spectrum.

%%=======================================================
\subsection{Modeling the HEG spectrum} \label{sec:predicted}
%%=======================================================

To calculate the predicted spectrum from the plasma surrounding \SgrA, we only include emission from regions physically probed by the HEG (described in Paper I). We also incorporate extinction from the interstellar medium (ISM). We calculate the predicted HEG spectrum for the +1 and -1 orders separately instead of combining them due to the differences in the instrumental response matrices, which would be more apparent in this low count regime. The background regions extracted from \citet{Corrales2020} were stacked by roll angle, with the background ``up'' or ``down'' regions (in the cross-dispersion direction) opposite from PWN G359.95-0.04 and IRS~13E,  chosen to minimize contamination. Since the simulated stellar wind plasma is not radially symmetric, we slice the annular region in half along galactic coordinates; Galactic North refers to regions where $b > 0$ and Galactic South refers to regions where $b < 0$. These two halves roughly correspond to the background regions considered. We use the spectrum calculated solely from the Galactic South as our background model, which ensures that we are capturing the plasma that is opposite in projection from PWN G359.95-0.04 and IRS~13E. %, with the specific make up as the background region in the data. 

The standard spectral extraction for HETG-S data captures X-ray events  within a 1.5\arcsec\ radius in the dispersion direction and 3\arcsec\ along the dispersion axis\footnote{https://cxc.harvard.edu/proposer/POG/html/chap8.html}. The source region, with a 1.5\arcsec\ radius, is fully encompassed within the projected HETG extraction area. However, the background region extends beyond 1.5\arcsec\ in the cross-dispersion direction. Accordingly, we introduce a scaling factor to the predicted background spectra denoting the area captured by HETG, which is the ratio of the background extraction region to the Galactic South half-annulus region described above ($ f_B = 0.3 $).

Following the procedure described in Section 3.2 of Paper I, we take into account the slitless nature of the HETGS by implementing a powerlaw background model. We include extinction due to the interstellar medium, employing the ISM abundance table of \citet{Wilms2000} and the dust extinction cross-sections from the \texttt{ISMdust} model \citep{Corrales2016}. We use the publicly available Python package \texttt{pyxsis}\footnote{https://github.com/eblur/pyxsis} to apply the instrumental response functions and perform a joint fit for the background models of the HEG~+1 and HEG~-1 spectra from \citet{Corrales2020}.

%%============================================
\section{Results} \label{sec:results}
%%============================================

We calculated line profiles from the two accretion modes (a disk versus a quasi-steady flow) that arise in the stellar wind simulations. We also compare the goodness-of-fit metrics for the stellar wind models  with the best fitting radiatively inefficient accretion flow (RIAF) model found in Paper I.

%%===================================================
\subsection{Searching For Signatures of a Cold Disk} \label{sec:results_a}
%%===================================================

There is evidence of a cold gas disk around \SgrA\ that simulations demonstrate could arise from shocked stellar winds. Observations of H30$\alpha$ emission indicate the presence of a cold disk of gas $\sim 10^4 \rm K$ that extends about 0.23\arcsec ($\sim 4.6 \times 10^4 R_g$) wide and is co-spatial within the hot accretion flow \citep{murchikova2019}. Interestingly, this observed cooler gas disk rotates counter to the hot plasma, but aligns with the rotation of the Galaxy, the circumnuclear disk, the mini-spiral, and the clockwise stellar disk \citep{murchikova2019}. We find that a cool disk naturally arises in the stellar wind plasma in both particle-based and grid-based simulations. However, in both cases, the simulated cold gas disk has an angular momentum vector offset by 90$^\circ$ orientation with respect to the observed disk. This misalignment suggests that other interacting structures, such as the minispiral or circumnuclear disk, which have not been incorporated into the simulations, may play a crucial role in its formation. This cold gas disk would not emit in the X-ray, but the presence of the cool disk heightens the accretion rate, resulting in slightly different plasma conditions (see Figure \ref{fig:sims}) which might be resolvable with X-ray spectroscopy. Differences in X-ray line profiles arising due to numerical method (i.e. particle-based vs. grid-based simulations) are investigated further in Calderón et al. 2024 (in prep).

\begin{figure*}
    \centering
    \includegraphics[width=0.365\textwidth]{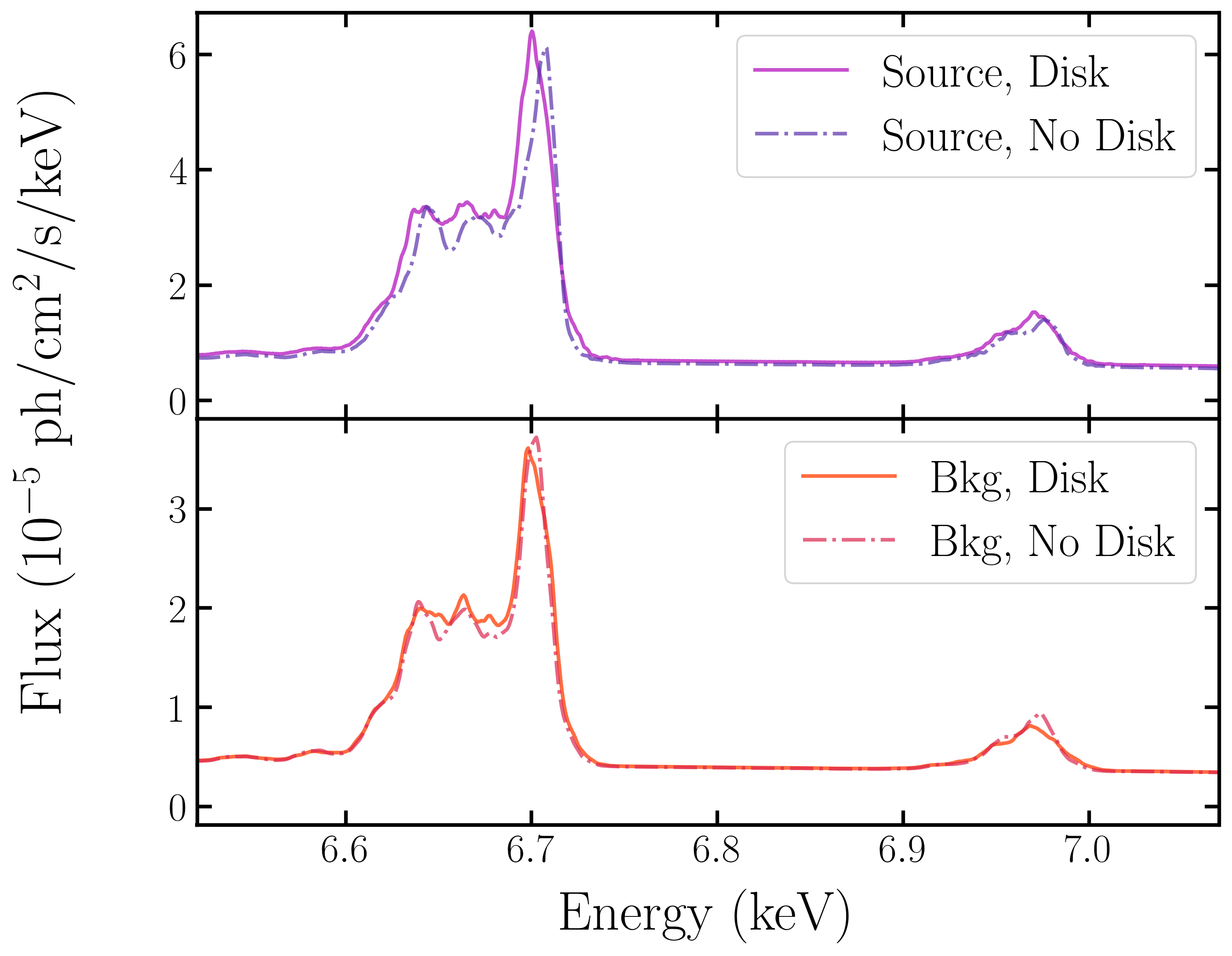}
    \hspace{0.015\textwidth}
    \includegraphics[width=0.594\textwidth]{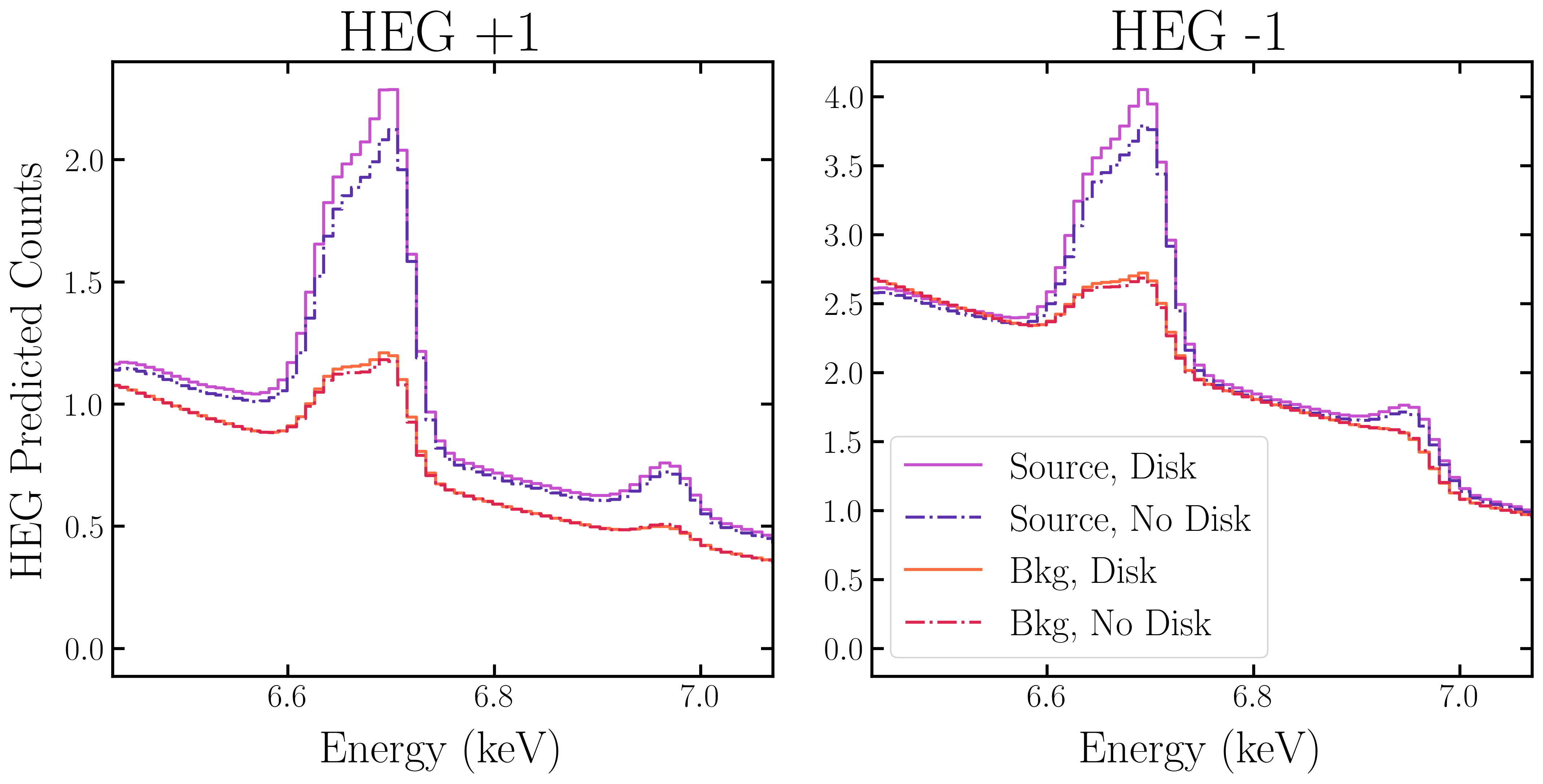}
    \caption{LEFT: Line profiles generated by plasma simulated in the 0\arcsec - 1.5\arcsec source region (top) and 1.5\arcsec - 5\arcsec  background region (bottom). %, in an updated use of procedure from R17. 
    %The y-axis is in units of $10^{-5} \rm ph/cm^2/s/keV$. 
    MIDDLE \& RIGHT: Predicted counts histograms calculated for HEG +1 (middle) and HEG -1 (right) for the disk (solid curves) and no-disk (dashed curves) plasma scenarios in the stellar winds simulation. We plot the source and background regions in purples and reds, respectively. The small differences between the final spectra with and without a cold gas disk are slight and not resolvable with \Chandra-HETG, differing only by $\sim$ 2.6\% at the peaks. We are thereby unable to detect evidence of the $\sim$$10^4$ K disk from X-ray spectroscopy alone.
    }
    \label{fig:stellar_wind_profiles}
\end{figure*}

To test whether we can resolve the differences between a hot plasma with and without a cool gas disk, we calculated the X-ray emission arising from both scenarios and generated predicted \Chandra-HETG spectra. Figure~\ref{fig:stellar_wind_profiles} (left) compares the simulated X-ray gas emission arising from a plasma with and without a cold gas disk. The top row pertains to plasma within the source extraction region, ranging from 0\arcsec\ to 1.5\arcsec, while the bottom row shows plasma within the Galactic South section of the background annulus, spanning from 1.5\arcsec\ to 5\arcsec. Within the plots, the disk and no-disk scenarios are represented by solid and dash-dotted lines, respectively. 
Unfortunately, the HEG predicted spectra from the two accretion modes are almost indistinguishable. %, as shown in the middle (HEG +1) and right (HEG -1) panels of Figure \ref{fig:stellar_wind_profiles}. 
The distinction between the disk and quasi-steady flow scenarios is clearest in the Fe K$\alpha$ emission features, where the peaks are shifted, but not to a degree that can be easily resolved by \Chandra. The maximum vertical difference between the two spectra is 2.6\%. Therefore, our current dataset is not sufficient to differentiate between the different plasma conditions that arise in the disk accretion and quasi-steady flow modes. Given the observational evidence of a cool accretion  disk surrounding \SgrA\ \citep{murchikova2019}, we only consider the X-ray line profiles from the disk scenario, moving forward.

%%===================================================
\subsection{Stellar Winds vs. RIAF} \label{sec:results_b}
%%===================================================
% other possible first sentence: We compare the predicted HEG spectra from the stellar winds to the best-fit model from Paper I (Balakrishnan et al. 2024a) in order to ascertain which model fits better. 
In Paper I, we used \texttt{pyatomdb} to calculate the emission from plasma in Coronal Ionization Equilibrium (CIE) governed by density and temperature profiles according to the Radiatively Inefficient Accretion Flow (RIAF) parametrization \citep{Yuan2003, Xu2006, Narayan2012}. The main parameters are the slope of the temperature profile, $q$, and the steepness of the mass accretion rate and electron density, $s$. We also fit for the relative abundance of Iron in \texttt{pyatomdb}, leading to a total of three free parameters in our RIAF model: $s$, $q$, and $Z_{Fe}$. The predicted HEG spectrum was computed with the same method as the stellar wind emission, accounting for geometry, instrumental effects, and ISM extinction. A significant distinction between the RIAF and stellar wind models is that the former does not take into consideration any potential velocity structure inside the plasma, including rotation. This has limited impact on our results, as extinction coupled with the HEG instrument response wash out the fine structure (see Paper I for more details). A Monte-Carlo Markov Chain (MCMC) fit to the data yields best-fit and 90\% credible intervals of $s = 0.74~(-0.12, +0.66)$, $q = 0.22~(-0.22, +1.35)$, and $Z_{Fe} = 0.13 Z_\odot~(-0.10, +0.19)$. The best-fit RIAF $s$ value is roughly consistent with an inflow balanced by an outflow ($s \sim 1$). The data prefer a RIAF model with a low Iron abundance, where $Z < 0.3 Z_\odot$.

Similarly, we use the publicly available \texttt{emcee} package to perform an MCMC fit for the background model with the static spectral model computed from the hydrodynamic simulations of WR stellar winds. 
The background contamination comes from different regions of the Galactic Center X-ray image, so the powerlaw parameters are allowed to be different for the +1 and -1 HEG orders, yielding a total of four free parameters. 

We fit the source and background regions for both orders simultaneously, using Cash statistics. We initiated the model with 20 walkers and applied uniform priors of  $-20 < \gamma < 20$ and $-12< \log{N_{PL}} < -3$. We ran the MCMC chains until the autocorrelation time, which measures the number of steps required for the chain to achieve a state independent of its previous state, was less than 5\%. The resulting best fits and  90\% credible intervals for the powerlaw components are: $ \log{N_{PL, +1}} = -7.6~(-0.1, +0.1)$, $ \gamma_{+1} = 6.7~(-6.7, +7.4) $, $  \log{N_{PL, -1}} = -7.2~(-0.1, +0.1)$, $ \gamma_{-1} = 4.1~(-3.8, +3.7) $. 

% Stellar Wind Params: 
% HPD for Norm (+1): [-7.70434562 -7.49334501]
% HPD for Gamma (+1): [ 0.04175308 14.1015002 ]
% HPD for Norm (-1): [-7.25180923 -7.14834574]
% HPD for Gamma (-1): [0.23290601 7.75874806]

% Best fit for Norm (+1): -7.58 (-0.08, +0.05)
% Best fit for Gamma (+1): 6.69 (-3.95, +4.53)
% Best fit for Norm (-1): -7.2 (-0.03, +0.03)
% Best fit for Gamma (-1): 4.05 (-2.2, +2.33)

\begin{figure*}[ht!]
    \centering
    \includegraphics[width=0.98\textwidth]{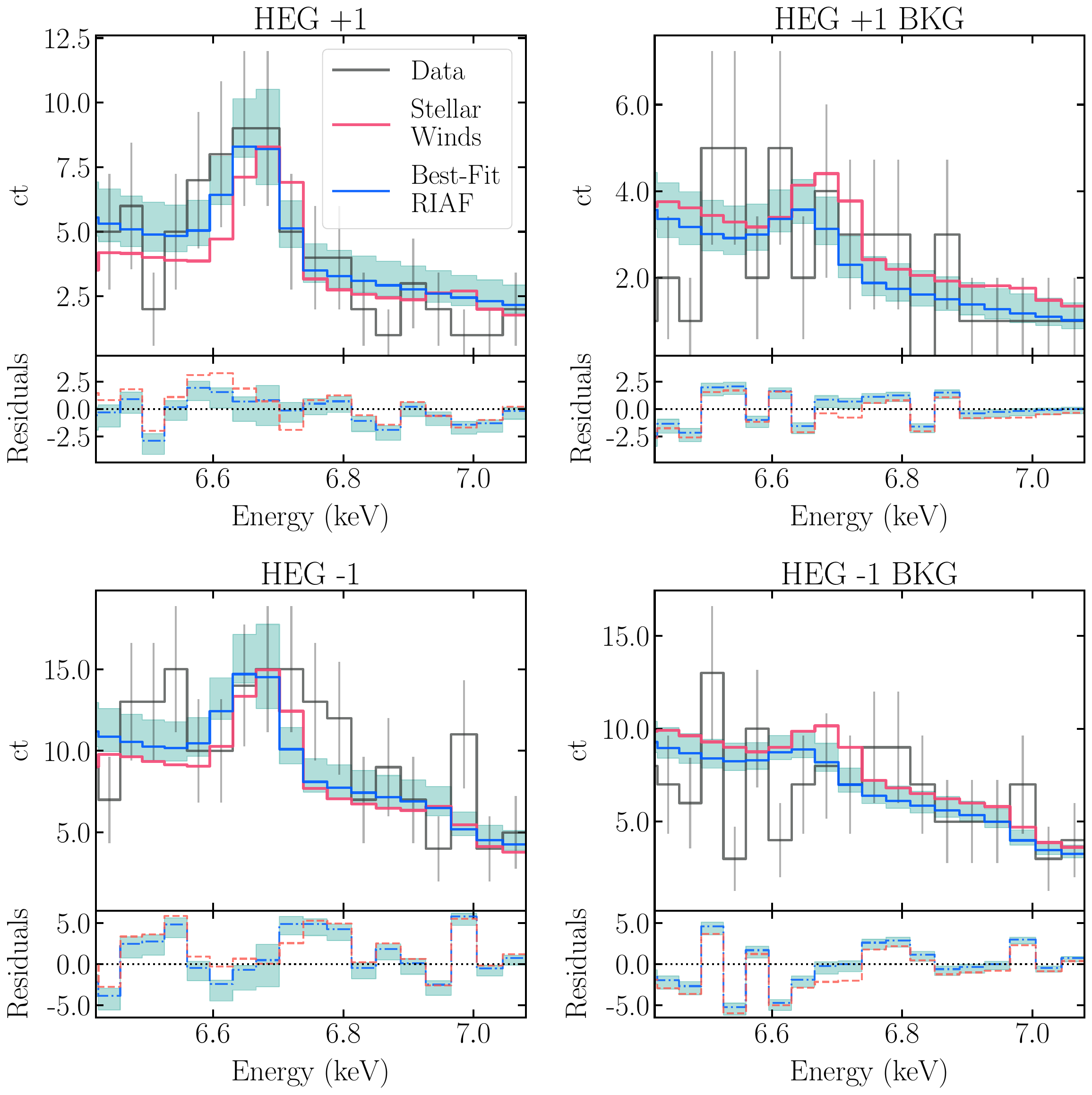}
    \caption{Final fits to the HEG +1 and -1 source and background regions. The observation is plotted in black, with the best-fit RIAF model in blue and the stellar wind simulations plotted in pink. The residual counts spectrum (observation $-$ model) are plotted below each subplot. The data is binned to a factor of 4 for visual clarity, however, the fitting was performed on the unbinnned HEG data. While the models capture the 6.7 keV peak in the source spectrum, most of the observed line structure is not predicted in the simulations. }
    \label{fig:finalfit}
\end{figure*}

In Figure \ref{fig:finalfit}, we plot the stellar wind plasma model on top of the best-fitting background model (pink), best-fit RIAF from Paper I (blue), and the observed HEG spectra (black). We have binned the models and data to a factor of 4 for visual clarity, including the residuals. Focusing solely on the model counts histograms, it is intriguing that the simulation of stellar winds and the RIAF model both predict peaks in Iron at 6.7 keV with comparable strengths. The line intensity in the simulations is determined by assuming a plasma abundance of $0.6-1~Z_\odot$ \citep{Asplund2009} depending on the specific WR subtype, whereas the RIAF fit yielded a lower value of $Z_{\rm Fe} = 0.13 Z_\odot$ \citep{AG1989}. Converting from \citet{AG1989} to \citep{Asplund2009} yields a $Z_{RIAF} = 0.18~Z_\odot$ and a 90\% credible upper limit of $0.45~Z_\odot$. %

We investigated whether differences in the temperature and density profiles from both models could explain the lower metallicity predicted by the RIAF model. The WR plasma model predicts elevated plasma temperatures compared to the RIAF temperature profile. At higher temperatures, a higher fraction of the Iron will be in the Fe XXVI state, which reduces the strength of the 6.7 keV line complex. Therefore, the cooler temperatures in the RIAF model are able to produce the same strength 6.7 keV line with less total Iron abundance. One major difference between the RIAF and WR spectral models is that the WR plasma predicts a more significant 6.97 keV feature from Fe XXVI. We are unable to detect this feature in the HEG spectrum due to the very low effective area at this energy.

Examining the fits to the data, both models capture some of the features in the HEG +1 and -1 source spectra. However, the HEG~-1 model spectra recreate neither the broadening in the 6.7 keV line nor the features at 6.5 and 7.0 keV. The background and extra features for the HEG~-1 are likely influenced by its placement on a back-illuminated chip, which has a higher background, as opposed to rotation or other broadening effects inherent in the plasma. Both the simulations and the RIAF model anticipate a 6.7 keV feature in the background area that is not very pronounced, suggesting that other sources of photons are likely more dominant $3 - 10 \times 10^6 R_g$ away from \SgrA. Indeed, the flux from the powerlaw contributes 95\% - 98\% of the flux in the background spectra.

\begin{figure}[ht!]
    \centering
    \includegraphics[width=0.48\textwidth]{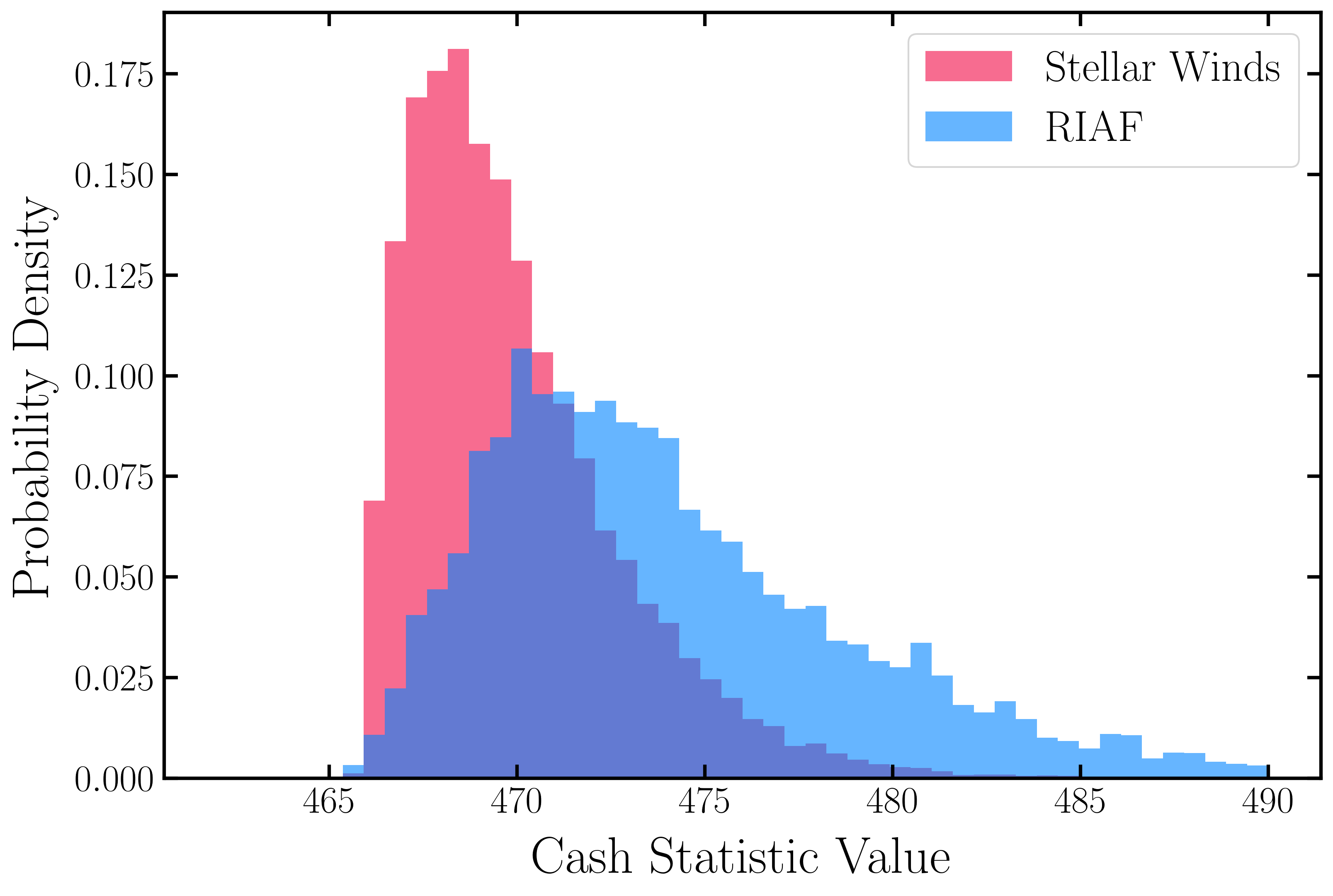}
    \caption{Histograms showing the Cash statistic from the posterior distributions of both the RIAF (blue) and stellar wind simulation fits (pink). }
    \label{fig:cashstat}
\end{figure}

In all procedures, the Cash statistic was used to evaluate goodness-of-fit in our MCMC sampling. Figure \ref{fig:cashstat} shows the distributions of the Cash statistic resulting from fitting the stellar wind plasma (pink) and the RIAF model (blue). While the stellar wind plasma model may appear to offer a better fit, there is considerable overlap in the Cash statistic distributions. To evaluate which model is statistically preferred, we utilize the Bayesian model evidence. 

Bayesian inference provides a consistent approach to the estimation of a set of parameters in a model $H$ using Bayes' theorem \citep{feoz_bayesian_2013}. For a helpful overview of Bayesian inference in astronomy and variation in notation, see \citet{eadie_bayesian2023}. Bayes theorem provides the probability distribution of model $H$ parameters, $\Theta$, given the dataset, $D$:
\begin{equation}
p(\Theta|D,H) = \frac{p(D|\Theta,H) p(\Theta|H)}{p(D|H)}.
\end{equation}
On the left, $p(\Theta|D,H)$, we have the posterior distribution. On the right, $p(D|\Theta, H)$ is the model likelihood $\mathcal{L}(\Theta)$, for which we use the Cash statistic, and $p(\Theta|H)$ is the prior for the model parameters. The denominator, $p(D|H)$, refers to the marginal likelihood or Bayes/model evidence. The Bayes evidence does not depend on the parameters, and is not used in parameter estimation, but is helpful when comparing two models. Selection between competing models $H_0$ and $H_1$ can be done by comparing their respective posterior probabilities to compute the model odds (Bayes Factor),  $\mathcal{R}$, as follows:
%% LC last reviewed here, 3/18/24

\[ \mathcal{R} = \frac{p(H_1|D)}{p(H_0|D)} = \frac{p(D|H_1)}{p(D|H_0)}\frac{p(H_1)}{p(H_0)} \]
%where we have used Bayes theorem,   
where $p(H_1) / p(H_0)$ is the prior probability ratio for the two models.  Once the posterior has been determined, the model evidence, $p(D|H)$ can be evaluated numerically by integrating the likelihood over the prior: $p(D|H) = \int L(\Theta) p(\Theta|H)  d^n(\Theta)$ over the parameter space. Calculating the Bayes evidence involves an integral that can be numerically challenging; several methods have arisen to address this \citep{feroz2009,handley2015,cai2021}.  We use the learnt harmonic mean estimator, which uses machine learning techniques to approximate the harmonic mean of posterior probabilities, providing a more efficient estimation method compared to traditional sampling \citep{McEwen2021}. By leveraging predictive models trained on observed data, it offers a more tractable computational solution for estimating the Bayesian evidence. We used the open source code \texttt{harmonic} \citep{McEwen2021} to compute the model evidence and the Bayes Factor from the posterior samples calculated in our \texttt{emcee} runs. 
%
%As the average of the likelihood over the prior, 
The Bayesian evidence is larger for a model if more of its parameter space has high likelihood values and smaller for a model with large areas in its parameter space having low likelihood values. %, even if the likelihood function is very highly peaked. 
Thus, using the Bayes evidence for model selection accounts for the differences in model complexity between our RIAF and stellar wind models. 
The value of the Bayes factor, $\mathcal{R}$, is often interpreted using Jeffreys' scale \citep{jeffreys_theory_1998, lee_bayesian_2013}, where $0 < \log{\mathcal{R}} < 2$, $2 < \log{\mathcal{R}} < 6$, and $6 < \log{\mathcal{R}} < 10$ indicate the ranges for weak, moderate, and strong evidence for model $H_1$ over model $H_0$, respectively. With $H_1$ denoting the RIAF model and  $H_0$ denoting the stellar wind simulations, we find that  $\log{\mathcal{R}} \sim 2$ , indicating a weak favour towards the RIAF model. However, the Bayes evidence is not strong enough to support either model for the accretion flow of \SgrA.  We note that non-linearities in the parameter space for the RIAF model make calculating the Bayes evidence challenging, adding to our conclusion the data do not strongly prefer either model.
\subsection{\SgrA\ at micro-calorimeter resolution}
\label{sec:microcal}

\begin{figure}[ht!]
    \centering
    \includegraphics[width=0.48\textwidth]{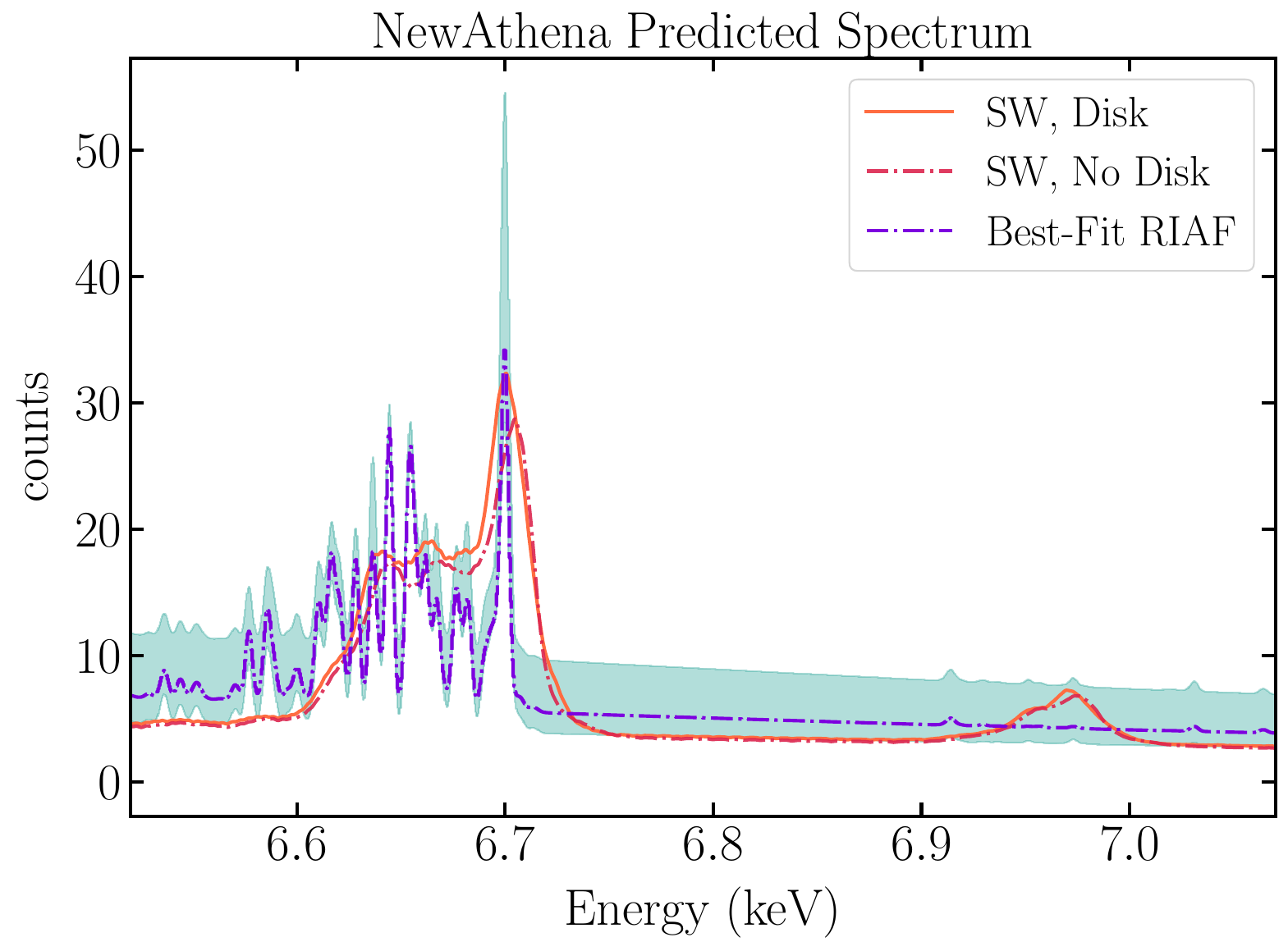}
    \caption{The anticipated \textit{NewAthena} photon counts spectra for the stellar wind plasma models (red) under both disk (solid) and quasi-steady (dot-dashed) accretion scenarios in the \FeKalpha\ region. The best-fit RIAF and contours from Paper I are overlaid in blue. The stellar wind model will be differentiable from the RIAF model, and it is possible, with a high signal-to-noise, to detect signatures of the cold disk in the X-ray plasma.}
    \label{fig:future}
\end{figure}
While the \Chandra-HETG is not able to strongly differentiate between the stellar wind plasma and a RIAF model, new and future missions, especially those utilizing microcalorimeters, could theoretically elucidate the mechanics at play. The HETG-S provides high spectral resolution for point sources, reaching $\sim$ 60-70 eV at 6.7 keV, coupled with a spatial resolution of 0.5\arcsec. The approved ESA L-class mission \textit{NewAthena}, launching late 2030s, will have an imaging resolution of $9$\arcsec, and the X-IFU microcalorimeter will provide a spectral resolution of $4-5$~eV at 6.7~keV.\footnote{\url{https://www.cosmos.esa.int/web/athena/supporting-sci-documents}} 
The advantages of microcalorimetry over a dispersive spectrograph can be seen in Figure \ref{fig:future}, where we have convolved our best-fit RIAF and stellar wind models out to 5\arcsec, including ISM extinction effects, with the \textit{NewAthena} X-IFU instrument response function. The model spectra were calculated with an exposure time of 100~ks.  Unfortunately, the spatial resolution of \textit{NewAthena} will result in spectral contamination from both PWN~G359.95-0.04 and IRS~13E, which lie a projected distance of approximately 4\arcsec\ from \SgrA. It is unlikely that the signal-to-noise will be high enough to differentiate between the disk and no-disk models, but it may be possible to distinguish between the simplified RIAF model and the predictions of hydrodynamic simulations. In particular, the heightened temperatures anticipated by the stellar wind simulations can be readily tested through the detection of the 6.97 keV Fe~XXVI line, which is not predicted by the current RIAF model.
The \textit{Lynx}\footnote{https://www.lynxobservatory.com/mission} concept mission is the only concept mission that would significantly advance our understanding of the X-ray emitting accretion flow of \SgrA. With a spatial resolution approaching 1\arcsec, \textit{Lynx} would be able to isolate \SgrA\ from nearby PWN G359.95-0.04 and IRS 13E. The unprecendeted sensitivity of \textit{Lynx} would open a completely new window on the \SgrA\ accretion flow, possibly allowing for constraints on the bulk and turbulent motions of the hot plasma.
%

%%============================================
\section{Summary \& Conclusions} \label{sec:conclusions}
%%============================================

Besides black hole accretion models, hydrodynamic simulations of shocked plasma from colliding winds of nearby WR stars alone can accurate reproduce the X-ray spectrum of \SgrA. We investigate whether the formation of a cold accretion disk, seen in both grid-based \citep{Calderon2020} and particle-based codes (this work), leaves signatures in the gas resolvable by X-ray spectroscopy. The quiescent high-resolution spectrum from the the legacy \Chandra-HETG dataset (PIs: Markoff, Nowak, \& Baganoff)\footnote{https://www.sgra-star.com/} may hint at complex velocity structure in the plasma \citep{Corrales2020}, with possible line shifts close to the maximum line-of-sight velocities predicted by the simulations ($v \sim 1500 \rm km \ s^{-1}$). However, this line structure could be due to extraction techniques or stochastic fluctuations from the relatively high X-ray background.

We compare the HETG-S spectrum to hydrodynamic models that incorporate complicated velocity structure in the dynamic plasma, accounting for the unique geometry and extraction processes of the \Chandra\ HETG-S instrument. % to get a deeper understanding of this spectrum. 
Building on the work of R17, we examine the X-ray emission from updated simulations of the plasma environment around \SgrA\ created by shocked WR stellar winds, using the smoothed particle hydrodynamics (SPH) code $\texttt{GADGET-2}$, and incorporating updates for the SMBH mass and stellar separations. These models do not consider SMBH feedback but include optically-thin radiative cooling. 

We find that the stellar wind plasma cools enough to form a cool disk within the hot accretion flow in both particle-based and grid-based simulations. We are able to reproduce a disk similar to the observed $\sim 10^4$~K H30$\alpha$ recombination disk \citep{murchikova2019} when we run our SPH simulations for 3500~yr. We generate column density, X-ray intensity, and velocity maps for plasma conditions with and without the cold disk. There are minor variations in the X-ray line profiles resulting from the different accretion scenarios, evidenced by slight shifts in the \FeKalpha\ peaks. However, these differences are not discernible with the current signal and spectral resolution of the \Chandra\ HETG-S.

We compare the predicted X-ray counts histogram from the SPH simulations to a general RIAF model, described in Paper I (Balakrishnan et al. 2024a). We find that the stellar winds make the plasma hotter than predicted by the RIAF.
%Neither model fully accounts for the observed line structure. 
Utilizing Bayesian model evidence to select between models, we find that the RIAF model is weakly preferred. Therefore, hints of a velocity-shifted line complex in the \Chandra\ high-resolution HETG spectrum is not proof of complex velocity and temperature structure, but is more likely due to statistical noise. Spectra of \SgrA\ taken by the future mission \textit{NewAthena} may be able to differentiate between the RIAF model and stellar wind emission, depending on the strength of the observed 6.97 keV Iron line. Identifying plasma signatures of the cold disk will still be difficult solely with X-ray data. If stellar winds simulations alone, which do not include a feedback mechanism launched from within $10^4 R_g$, sufficiently explain the quiescent X-ray spectrum of \SgrA, it suggests that accretion stalls in the outer regions of the accretion flow and that feedback is not limited to regions close to the event horizon. This implies that the day-to-day flares from \SgrA\ are not necessarily the main drivers of feedback. 

While \Chandra\ offers the best current combination of spatial and spectral resolution, of all future X-ray missions proposed, only \textit{Lynx} would be able to isolate the Sgr A* accretion flow from its surroundings and provide high enough sensitivity to make significant advances constraining the structure of the \SgrA\ accretion flow between $10^4$ and $10^6 R_g$.

\begin{acknowledgments}
We would like to thank the referee for their insightful responses, which greatly improved our manuscript. MB thanks Kayhan Gultekin and Kaze Wong for insight into model comparison statistics. JC thanks Alex Gormaz-Matamala for his pointers to the WR abundance literature. The original \Chandra\ dataset used in this work was made possible by the Chandra X-ray Visionary Program through Chandra Award Number GO2-13110A, issued by the Chandra X-ray Center (CXC), which is operated by the Smithsonian Astrophysical Observatory for NASA under contract NAS8-03060. Authors MB, CR, and LC were supported by the CXC grants program, award AR2-23015A and AR2-23015B. DC is funded by the Deutsche Forschungsgemeinschaft (DFG, German Research Foundation) under Germany’s Excellence Strategy – EXC 2121 – ‘Quantum Universe’ - 390833306. JC acknowledges financial support from ANID (FONDECYT 1211429 and Millenium Nucleus TITANS, NCN2023\_002). DH acknowledges funding from the Natural Sciences and Engineering Research Council of Canada (NSERC) and the Canada Research Chairs (CRC) program. Resources supporting this work were provided by the NASA High-End Computing (HEC) Program through the NASA Advanced Supercomputing (NAS) Division at Ames Research Center.
\end{acknowledgments}

\software{matplotlib \citep{Hunter:2007}, astropy \citep{2013A&A...558A..33A,2018AJ....156..123A}, pyatomdb \citep{pyatomdb}, pyxsis \footnote{https://github.com/eblur/pyxsis}, harmonic \citep{McEwen2021}, gadget-2 \citep{Springel2005}
          }



% A handy "cheat sheet" that provides the necessary \latex\ to produce 17 
% different types of tables is available at \url{http://journals.aas.org/authors/aastex/aasguide.html#table_cheat_sheet}.

\bibliography{refs}{}
\bibliographystyle{aasjournal}



\end{document}